\documentclass[journal]{IEEEtran}
\usepackage{graphicx}

\begin{document}
\title{The CALICE Software Framework and Operational Experience}
\author{Mark~Terwort,
  Angela~Lucaci-Timoce,~\IEEEmembership{Fellow,~CERN}, for the CALICE
  collaboration
  \thanks{Mark Terwort is with DESY Hamburg, Germany (e-mail: mark.terwort@desy.de).}
  \thanks{Angela Lucaci-Timoce is a fellow in the Linear Collider
    Group at CERN, e-mail:
    angela.isabela.lucaci.timoce@cern.ch.}}

\maketitle
\pagestyle{empty}
\thispagestyle{empty}

\begin{abstract}
  The CALICE collaboration is developing calorimeters for a future
  linear collider, and has collected a large amount of physics data
  during test beam efforts. For the analysis of these data, standard
  software available for linear collider detector studies is applied.
  This software provides reconstruction of raw data, simulation,
  digitization and data management, which is based on grid tools. The
  data format for analysis is compatible with the general linear
  collider software. Moreover, existing frameworks such as Marlin are
  employed for the CALICE software needs. The structure and features
  of the software framework are reported here as well as results from
  the application of this software to test beam data.
\end{abstract}


\section{Introduction}

\IEEEPARstart{T}{he} CALICE collaboration~\cite{CALICE} is formed of
more than 300 physicists and engineers from Europe, America, Asia and
Africa. Its purpose is to carry on a research and development program
of hadronic and electromagnetic calorimeters for a future Linear
Collider at the TeV scale.

Several test beam campaigns (see Tab.~\ref{tab:Calice-tbeams}) were
successfully performed by the collaboration with different
combinations of detectors: a hadronic calorimeter (HCAL) using steel
or tungsten as absorber and scintillator tiles, read out by silicon
photomultipliers (SiPMs), as active material~\cite{Adloff:2010hb}, a
Si-W ECAL~\cite{Anduze:2008hq}, a Tail Catcher and a digital HCAL
based on gas proportional chambers (RPCs)~\cite{Bilki:2009wp}.

The detectors were accompanied by DAQ systems, triggers and drift
chambers, which provide tracking information. The test beam campaigns
resulted in several tens of terabytes of data saved on tape. For
lasting physics results, it has to be insured that the data are
treated consistently, despite the different types of detectors, and
their different states of development.

In the following, the efforts done by the CALICE collaboration in
developing the software needed to extract relevant physics results are
presented.

\begin{table}[h]
\begin{center}
\caption{Test beams carried on by the CALICE collaboration}
\label{tab:Calice-tbeams}
\begin{tabular}{|c|c|c|}
\hline  Location & Year & Detectors\\\hline
\hline DESY & 2006 & Scintillator/steel HCAL\\ 
\hline  CERN & 2007 & Si-W ECAL, \\
                    &           & scintillator/steel HCAL, \\
                     &           &Tail Catcher\\ 
\hline FNAL & 2007/2008 & Si-W ECAL, \\
                    &                   & scintillator/steel HCAL, \\
                    &                   & Tail Catcher\\ 
\hline FNAL & 2009 & Scintillator ECAL, \\
                    &          & scintillator/steel HCAL, \\
                     &         & Tail Catcher\\
\hline CERN & 2010 & Scintillator/tungsten HCAL \\
\hline FNAL & 2010 & Digital HCAL\\
\hline
\end{tabular} 
\end{center}
\end{table}

\section{The CALICE Data Flow}

At the experimental site, the data are recorded in binary format and
then transferred directly to the grid storage elements at DESY
Hamburg, which provides a tape back-up, and replicated, for safety
reasons, at the Computing Centre of IN2P3 at Lyon. The event building
is also done on the grid and generates files in the LCIO format (see
section~\ref{sec:structure}).  Since the year 2005, a total of about
50 TByte of raw and processed data and simulation files have been
managed using grid, i.e \textit{lcg} tools~\cite{lcg}.

In order to be analysed, the data need to be calibrated first. The
calibration depends on the detector type and usually implies the usage
of so-called calibration constants, which are extracted offline.

For a common storage of all calibration constants, the mappings of the
different channels and alignment information, a MySQL data base,
hosted at DESY, is employed by CALICE. Some of the information is
written into the data base during conversion and some, like
calibration constants, at a later stage by experts. Once the best
available information is written to the data base, the corresponding
folders are tagged. This ensures that the results can be reproduced
and cross-checked later without problems.

The access to the data base is done based on IP-ranges. As soon as a
new group joins the collaboration, their IP-ranges are added to the
list. Even if the access to the data base may be sometimes slow from
remote locations (like Japan), there is always the possibility of
dumping the data base information on files which can be stored
locally.

Actually, there are two instances of the data base: one for reading,
available for everybody, and one for writing and reading, for experts
work. The experts need to provide a password in order to be able to
modify the contents of the data base. For test purposes, user folders
are created. Once the expert is satisfied with the results, the
information is copied to the central folders.

\begin{figure}[hbt]
 \begin{center}
        \includegraphics[scale=0.32]{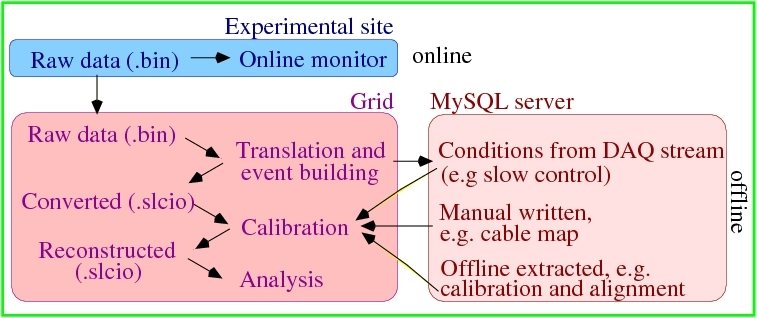}
 \end{center}
\caption{Schematic description of the CALICE data flow.}
\label{fig:calice-data-flow}
\end{figure}

A schematic representation of the CALICE data flow is shown in
Fig.~\ref{fig:calice-data-flow}.

The mass reconstruction is done centrally, upon request. This is based
on jobs submitted to the grid (involving centres from all over
Europe), as well as local batch farms.  Once the data is calibrated
(i.e. reconstructed), it can be analysed.

\section{The Structure of the CALICE Software}
\label{sec:structure}

The CALICE software is based on the ILC software~\cite{ILCsoft}. It
uses the C++ programming language, and the \textit{cmake}~\cite{cmake}
tool for creating platform independent Makefiles. The documentation is
done mainly inside the code, using \textit{doxygen}~\cite{doxygen}.

The development is done by people from the various groups. There is a
designated responsible for each group, and they are coordinated by a
central software person.

The software is maintained with an SVN server hosted by
DESY~\cite{svn}, and is organised in packages:
\begin{itemize}
\item {\it calice\_userlib} Contains general purpose classes, used in
  the other packages
\item {\it calice\_reco} The main package, contains the reconstruction
  code for the scintillator HCAL, Si-W ECAL and for the Tail Catcher
\item {\it calice\_lcioconv} Does the conversion from binary to LCIO
  format
\item {\it calice\_sim} Includes digitisation of simulated events
\item {\it calice\_run} Contains bash scripts for automatic generation
  of steering files, used for reconstruction, noise extraction and for
  the digitisation
\item {\it calice\_torso} Contains $HelloWorldProcessor$, as start-up
  for new users
\end{itemize}

In addition, an external package, called \textit{RootTreeWriter}, is
used to create ROOT trees for simple analyses.

The software releases are announced on the main web
page~\cite{CALICE-sw} and on dedicated mailing lists. With every
release, a tar-ball is created and installed on the grid, for
subsequent usage.

The simulation is realised by Mokka~\cite{Mokka}, which provides the
geometry interface to the GEANT4~\cite{GEANT4} simulation toolkit and
is also used for full detector simulation studies.  In order to save
time, the digitisation step, performed in the \textit{calice\_sim}
package is separated from the time consuming simulation step. The
simulation and digitisation of the data runs are done centrally, upon
request.

\subsection{CALICE Event Displays}
\begin{figure}[t!]
 \begin{center}
        \includegraphics[width=3.5in]{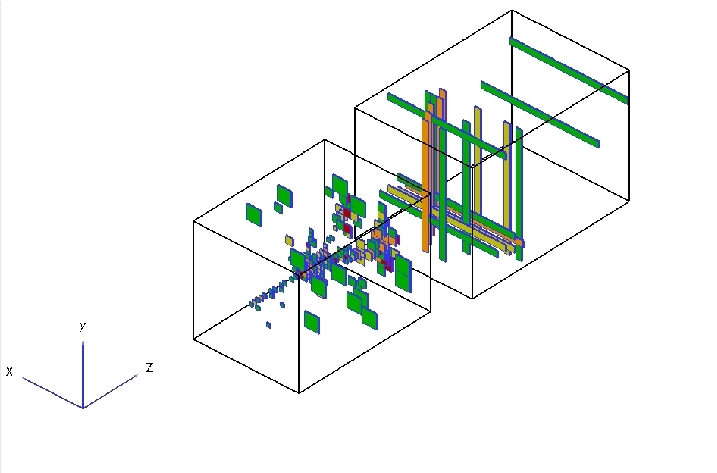}
 \end{center}
 \caption{Example of a CED based event display of a pion shower of 20
   GeV in the CALICE HCAL and Tail Catcher.}
\label{fig:eventDisplay}
\end{figure}

For displaying CALICE test beam events, currently two displays can be
used: one based on CED, which is the standard ILC event display, used
also for the full detector, and one based on ROOT geometry classes,
named DRUID~\cite{DRUID}. An example of such an event display is shown
in Fig.~\ref{fig:eventDisplay}.

\subsection{Testing}
Before each release, it is tested if the software compiles, and if it
produces the expected results. Nevertheless, since mistakes can always
happen, a better solution is to do the testing in an automatised way,
and to be able to easily compare results obtained with previous
software tags. This is done for the CALICE software using
\textit{ctest}~\cite{ctest}, which is a tool coming for free with
\textit{cmake}. This tool can be used for automatic updating from SVN,
for configuring, building, testing, memory checking and for submitting
results to a \textit{CDash} dashboard system. CALICE uses the
\textit{CDash} server installed by the ILC software group at DESY
Hamburg. Apart from automatisation, this has the advantage that the
outcome of the tests is stored in a central place, and the view of the
history in time is possible.

First basic {\it ctest} scripts are already being used for several
CALICE packages (\textit{calice\_userlib}, \textit{calice\_reco} and
\mbox{\textit{calice\_sim}}).

\section{Application of the CALICE Software - PandoraPFA}

In the beginning years of the collaboration, the software was written
in view of the immediate needs of the specific group. As the groups
evolved, more and more accent was put on modularity and flexibility,
since new coming detectors need to be integrated easily. In addition,
CALICE profits from the close collaboration with the ILC core software
developers. The advantages of this integrated strategy are underlined
by a recent analysis in which data recorded in test beam were subject
to an analysis using tools developed for the full detector
studies~\cite{Oleg}, namely the Pandora Particle Flow Algorithm
(PFA)~\cite{Thomson:2009rp}.

The aim of building a very high granular calorimeter is the capability
to measure the details of hadron showers and ultimately recover
neutral hadron energies in the vicinity of charged hadrons. This leads
to an increased overall jet energy resolution ($\sigma_E/E\sim
30\%\sqrt(E(GeV))$), since the energy of charged particles can be
measured in the tracking detectors with much higher resolution than in
the calorimeters. The resolution is degraded by the false assignment
of hits to overlapping particle showers from charged and neutral
particles. This depends on the deposited energy in the calorimeter as
well as on the distance of the showers.

To study these effects test beam data samples with charged pions with
energies from 10 to 50\,GeV have been used, which were taken during
the CERN test beam runs in 2007. The aim is to study the dependence of
the energy recovery capability of the PFA in events containing two
pions of the pion energies and their distance. The events containing
two pions have been constructed by overlaying two single pion events,
of which the energies have been measured with the standard procedure
in the calorimeter, and varying their energy and distance of the
shower axes. Additionally it has been assumed that one of the pions is
neutral and it has been studied how well the energy of this pion is
recovered by the PFA, after having mapped the event topology to a
Linear Collider detector geometry.

\begin{figure}[t!]
  \begin{center}
        \includegraphics[width=3.5in]{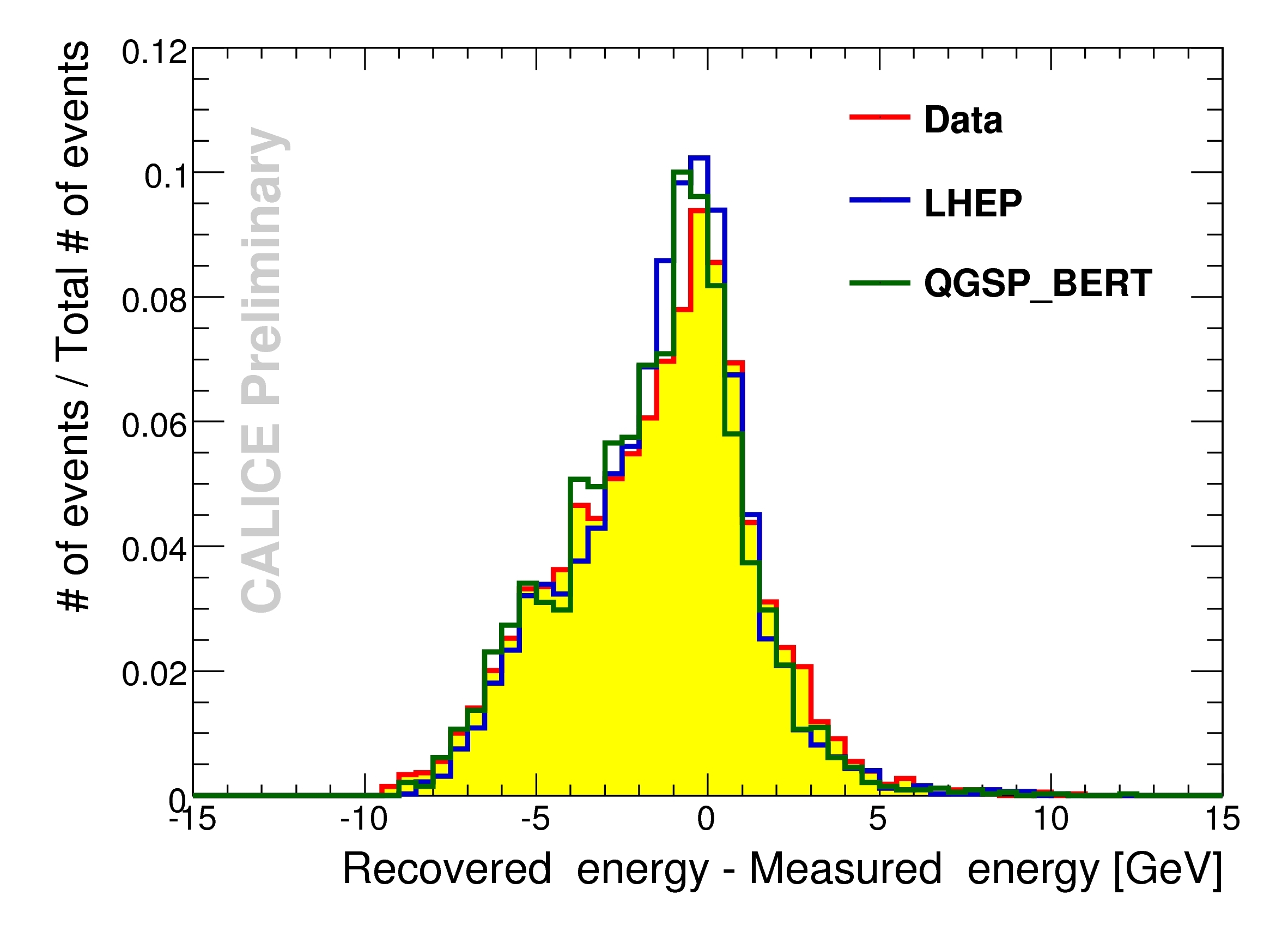}
 \end{center}
 \caption{Difference between the recovered energy and the measured
   energy for the 10\,GeV neutral hadron at 5\,cm distance from the
   10\,GeV charged hadron.}
\label{fig:PFA1}
\end{figure}

\begin{figure}[t!]
 \begin{center}
        \includegraphics[width=3.5in]{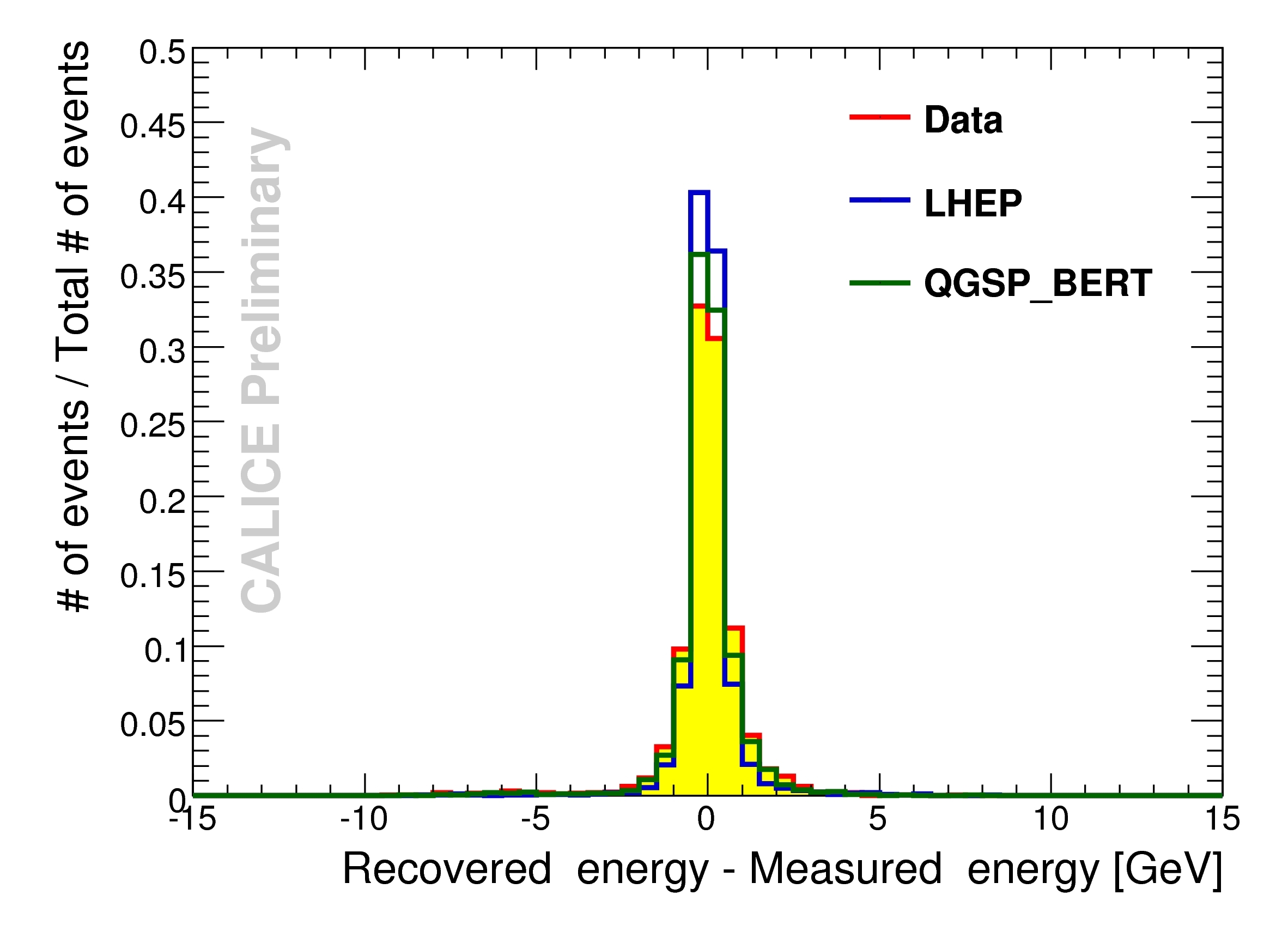}
 \end{center}
\caption{Difference between the recovered energy and the measured
   energy for the 10\,GeV neutral hadron at 30\,cm distance from the
   10\,GeV charged hadron.}
\label{fig:PFA2}
\end{figure}

\begin{figure}[t!]
 \begin{center}
        \includegraphics[width=3.5in]{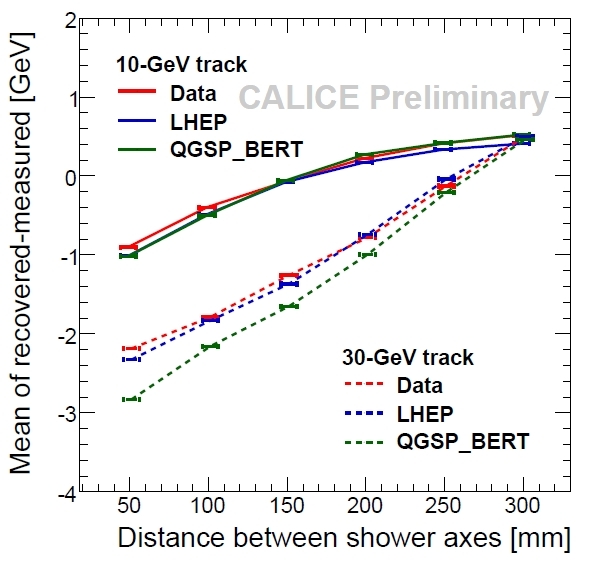}
 \end{center}
 \caption{Mean difference between the recovered energy and the
   measured energy for the 10\,GeV neutral hadron as a function of the
   distance to the 10\,GeV (solid) and 30\,GeV (dashed) charged
   hadron.}
\label{fig:PFA_summary}
\end{figure}

Fig.~\ref{fig:PFA1} shows the difference between the recovered energy
and the previously measured energy of a 10\,GeV neutral pion in the
vicinity of a 10\,GeV charged pion with a distance of 5\,cm. The
distribution is very broad and degrades even further for higher
charged pion energies. As a comparison Fig.~\ref{fig:PFA2} shows the
difference between the recovered energy and the previously measured
energy for pions of the same energy, but with a distance of 30\,cm.
The width of the distribution is considerably smaller, which shows
that the expected behaviour is reproduced by the algorithm. In both
plots not only the test beam data distribution is shown, but also the
MC predictions of the LHEP model and the QGSB\_BERT model. Apparently,
the simulation reproduces the results of the test beam data very well.
A summary plot showing the mean of the difference between the
recovered energy and the measured energy as a function of the distance
of the shower axes for two different charged pion energies is given in
Fig.~\ref{fig:PFA_summary}. As discussed, the confusion depends on the
radial distance between the showers, i.e. the showers overlap more at
smaller distances. Again, good agreement between data and MC
predictions is visible, which shows that the PandoraPFA is a reliable
reconstruction program for a full size detecctor.

\section{Conclusions and Outlook}

The CALICE collaboration has operated several test beam campaigns over
the last years and the analysis of the data requires powerful software
tools. This incorporates the data reconstruction and analysis as well
as the data management. The CALICE software makes use of the
developments done for the ILC analysis software and is fully
integrated into this framework. In particular Marlin processors are
used for the analysis and Mokka as the simulation framework. The
worldwide computing grid is heavily used both for data storage and
data processing. The CALICE software has been shown to scale for large
data sets during years of test beam data analysis. A study of the
PandoraPFA algorithm has been presented as an example of the
successful application of the CALICE software in data analysis and it
shows that it provides a reliable reconstruction for a full size
experiment.

The next step for the development of the CALICE software is the
integration of the technological prototypes as well as the Digital
HCAL (DHCAL) and the Semi-Digital HCAL (SDHCAL). Integration means for
example the usage of the LCIO data format and the common CALICE data
base. This is necessary, since the DHCAL started taking test beam data
recently and the SDHCAL will start data taking in 2011. Furthermore,
the second generation of the CALICE DAQ is currently under development
and has to be fully integrated into the CALICE software.

\section*{Acknowledgment}

The authors gratefully thank the whole CALICE collaboration for the
successful development and operation of the software framework as well
as for useful discussions and contributions to the results presented
here.

The software development has profited a lot from the close
collaboration with the ILC core software developpers, who offer the
framework, and continuous support. In addition, the data management is
strongly supported by the DESY grid groups.

\end{document}